%% file: virtual.tex
\documentclass[english,11pt]{article}
\usepackage[papersize={8.5in,11in},margin=1in]{geometry}
\input{header.tex}

\begin{document}


\let\scriptstyle\textstyle
\newcommand{\TL}{\fontsize{12}{18}}

\newcommand{\st}{\textrm{s.t.}}
\newcommand{\tld}{\tilde}
\newcommand{\crt}{\color{red}}

\title{\TL Generalizing Virtual Values to Multidimensional Auctions:  \\
a Non-Myersonian Approach\thanks{We thank Jieming Mao, Vahab Mirrokni,
Renato Paes Leme, and Balasubramanian Sivan for their helpful comments and
advice on an earlier version of the manuscript.}}
\author{Song Zuo\thanks{Institute for Interdisciplinary Information Sciences,
Tsinghua University, Beijing, China. \url{songzuo.z@gmail.com}.}}
\date{\today}

\maketitle

\begin{abstract}
  We consider the revenue maximization problem of a monopolist via a
  non-Myersonian approach that could generalize to multiple items and multiple
  buyers. Although such an approach does not lead to any closed-form solution of
  the problem, it does provide some insights to this problem from different
  angles. In particular, we consider both Bayesian (Bayesian Incentive
  Compatible + Bayesian Individually Rational) and Dominant-Strategy
  (Dominant-Strategy Incentive Compatible + ex-post Individually Rational)
  implementations, where all the buyers have additive valuations and
  quasi-linear utilities and all the valuations are independent across buyers
  (not necessarily independent across items).

  The main technique of our approach is to formulate the problem as an LP
  (probably with exponential size) and apply primal-dual analysis. We observe
  that any optimal solution of the dual program naturally defines the {\em
  virtual value functions} for the primal revenue maximization problem in the
  sense that any revenue maximizing auction must be a virtual welfare maximizer
  (cf. Myerson's auction for single item~\cite{myerson1981optimal}).\footnote{It
  is in fact implied by \cite{cai2012optimal}, while our results stand on the
  exact form of the virtual value functions.}

  Based on this observation, we have the following results (most of them are
  previously unknown for the multi-item multi-buyer setting):
  \begin{enumerate}
    \item We characterize a sufficient and necessary condition for BIC $=$ DSIC,
          i.e., the optimal revenue of Bayesian implementations equals to the
          optimal revenue of dominant-strategy implementations ($\BRev =
          \DRev$). The condition is if and only if {\em the optimal DSIC revenue
          $\DRev$ can be achieved by a DSIC and ex-post IR virtual welfare
          maximizer with buyer-independent virtual value functions} (buyer $i$'s
          virtual value is independent with other buyers' valuations).
    \item In light of the characterization, we further show that when all the
          valuations are i.i.d., it is further equivalent to that
          separate-selling is optimal. In particular, it respects one result
          from the recent breakthrough work on the exact optimal solutions in
          the multi-item multi-buyer setting by~\cite{yao2016solutions}.
    \item We also observe that dual programs can be interpreted as the optimal
          transport problem. This result is previously shown by
          \cite{daskalakis2013mechanism,daskalakis2015strong} for the single
          buyer setting. Thus we automatically obtain a generalized version for
          the multi-buyer setting.\footnote{For the multi-buyer setting, the
          dual program is a generalized version of optimal transport problem.}
    \item We provide an alternative proof of Myerson's auction. In particular,
          we can directly start with solving the optimal DSIC and ex-post IR
          auction instead of first solving the optimal Bayesian implementation
          then showing that it is also a dominant-strategy implementation.
  \end{enumerate}
\end{abstract}

\clearpage
\section{Introduction}

\paragraph{Roadmap}
We introduce our notations and the common definitions in \autoref{sec:notation}.
We apply the dual analysis and define virtual values in \autoref{sec:dual}. In
\autoref{sec:charac} we present our main characterization and we then show that
in \autoref{sec:iid} for the i.i.d. setting, such a charaterization implies that
DSIC $=$ BIC if and only if separate-selling is optimal.

\section{Notations}\label{sec:notation}

  Throughout this manuscript, we use subscripts ${}_{i}$ to indicate buyers
  and superscripts ${}\sci{j}$ to indicate items. We also use boldface (without
  subscript) notations for vectors across all the buyers (e.g., matrices for
  allocation $\x$ and value $\v$ while vectors for payments $\p$) and notations
  without superscripts (but with subscripts) for vectors across all the items
  for some certain buyer (e.g., allocation $x_{i}$ and value $v_{i}$ of
  buyer $i$, while both are vectors). As a general convention, we use subscripts
  ${}_{-i}$ for the vectors without the element(s) for buyer $i$ and $[n] =
  {1, \ldots, n}$ for the set of buyers and $[m] = {1, \ldots, m}$ for the set
  of items. We also use $\cdot$ to emphasize the inner product of between two
  vectors.

  As we will consider both Bayesian and dominant-strategy implementations, we
  use variables with $~\bar{}~$ for Bayesian implementations (e.g., $\bar
  x_{i}(\v)$ and $\bar p_{i}(\v)$) while those without $~\bar{}~$ for
  dominant-strategy implementations (e.g., $x_{i}(\v)$ and $p_{i}(\v)$).

  We consider the case where the buyers have independent values with each other
  (yet the values of the same buyer for different items might be correlated).
  The values are additive and the utilities are quasi-linear. We will formalize
  the definitions later.

  For ease of using linear programs, we consider discrete distributions with
  finite supports.\footnote{Generalization to arbitrary distribution (if
  possible) would require linear programs for infinite dimensions.}

  \subsection{Direct Auctions}

    A direct auction $M = \langle \x, \p \rangle$ consists of the allocation
    $\x: \R_+^{n \times m} \rightarrow [0, 1]^{n \times m}$ and the payment $\p:
    \R_+^{n \times m} \rightarrow \R_+^{n}$. The utility of each buyer $i$ is
    \begin{align*}
      u_{i}(\v) = v_{i} \cdot x_{i}(\v) - p_{i}(\v)
                   = \sum_{j \in [m]} v\scr{i}{j}x\scr{i}{j}(\v) - p_{i}(\v).
    \end{align*}

    For any value profile $\v \in \R_+^{n \times m}$, the allocation and payment
    must satisfy the following feasibility constraint, $\forall j \in [m],\v \in
    \R_+^{n \times m}$,
    \begin{align*}
      \One \cdot \x\sci{j}(\v) = \sum_{i \in [n]} x\scr{i}{j}(\v) \leq 1.
    \end{align*}

    We use $\mu$ to denote the probability measure of the common prior knowledge
    on the private values. In particular, since the values are independent
    across buyers, $\mu(\v)$ can be written as
    \begin{align*}
      \mu(\v) =
        \mu_{1}(v_{1}) \mu_{2}(v_{2}) \cdots \mu_{n}(v_{n}).
    \end{align*}

    Let $\V_{i} \subseteq R^m_+$ be any finite support of the prior
    distribution of buyer $i$, namely,
    \begin{align*}
      \forall v_{i} \in \V_{i},~\mu_{i}(v_{i}) \geq 0,~\text{and}~
      \sum_{v_{i} \in \V_{i}} \mu_{i}(v_{i}) = 1.
    \end{align*}

    Write $\V = \V_{1} \times \V_{2} \times \cdots \times \V_{n}$. In
    particular, we will assume that $\forall v_{i} \neq \Zero$,
    $\mu_{i}(v_{i}) > 0$ to simplify the discussion of corner cases.

  \subsection{Bayesian Implementation}

    A direct mechanism $\bar M = \langle \bar \x, \bar \p
    \rangle$ is Bayesian Incentive Compatible (BIC), if $\forall v_{i},
    v'_{i} \in \R^m_+$,
    \begin{align}\label{def:bic}\tag{\textrm{BIC}}
      \E_{\v_{-i}}\left[v_{i} \cdot \bar x_{i}(\v)
            - \bar p_{i}(\v)\right] \geq
        \E_{\v_{-i}}\left[v_{i}
              \cdot \bar x_{i}(v'_{i}, \v_{-i})
              - \bar p_{i}(v'_{i}, \v_{-i})\right];
    \end{align}
    Bayesian Individually Rational (BIR), if $\forall v_{i} \in \R^m_+$,
    \begin{align}\label{def:bir}\tag{\textrm{BIR}}
      \E_{\v_{-i}}\left[v_{i} \cdot \bar x_{i}(\v)
            - \bar p_{i}(\v)\right] \geq 0.
    \end{align}

    By restricting to the support space $\V$, we can define the optimal Bayesian
    direct mechanism as the following linear program:
    \begin{align}
      \max \quad & \sum_{\v} \mu(\v) \sum_i \bar p_{i}(\v)
                   \label{lp:bayes}\tag{\textrm{BLP}}  \\
      \st  \quad & \sum_{\v_{-i}}\mu_{-i}(\v_{-i})(
                      v_{i} \cdot \bar x_{i}(\v) - \bar p_{i}(\v))
                   \geq \sum_{\v_{-i}}\mu_{-i}(\v_{-i})(
                      v_{i} \cdot \bar x_{i}(v'_{i}, \v_{-i})
                        - \bar p_{i}(v'_{i}, \v_{-i})),
                   ~\forall i \in [n],~v_{i}, v'_{i} \in \V_{i}
                   \nonumber  \\
                 & \sum_{\v_{-i}}\mu_{-i}(\v_{-i})(
                      v_{i} \cdot \bar x_{i}(\v) - \bar p_{i}(\v))
                   \geq 0,~\forall i \in [n],~v_{i} \in \V_{i}
                   \nonumber  \\
                 & \sum_i \bar x\scr{i}{j}(\v) \leq 1,
                   ~\forall j \in [m],~\v \in \V \nonumber  \\
                 & \bar x\scr{i}{j}(\v), \bar p_{i}(\v) \geq 0,
                   ~\forall i \in [n],~j \in [m],~\v \in \V \nonumber
    \end{align}

    Although any feasible solution to this linear program only defines the
    allocation and payments for the value profiles in the support space and the
    \ref{def:bic} and \ref{def:bir} properties are only guaranteed within the
    support space, there is a standard extension method to recover the Bayesian
    implementation that is (i) defined on the full value space
    $\R_+^{n \times m}$, (ii) \ref{def:bic} and \ref{def:bir} (on the full value
    space), and (iii) consistent with the given feasible solution on support
    space $\V$.

    \begin{lemma}[Bayesian Extension]
      Given any feasible solution $(\bar \x, \bar \p)$ to \ref{lp:bayes}, the
      extended direct mechanism $(\bar \x', \bar \p')$ defined as follows
      satisfies \ref{def:bic} and \ref{def:bir}.
      \begin{align*}
        &~\bar \x'(\v) = \bar \x(\v'),~ \bar \p'(\v) = \bar \p(\v')  \\
        \text{where}~\forall i \in [n],~&~\text{if}~v_i \in \V_i,~v'_i = v_i  \\
        &~\text{otherwise},~v'_i = \argmax_{v_i \in \V_i}
            \E_{\v_{-i}}\left[v'_{i}
              \cdot \bar x_{i}(v_{i}, \v_{-i})
              - \bar p_{i}(v_{i}, \v_{-i})\right].
      \end{align*}
    \end{lemma}

  \subsection{Dominant-Strategy Implementation}

    A direct mechanism $M = \langle \x, \p \rangle$ is Dominant-Strategy
    Incentive Compatible (DSIC), if $\forall \v \in \R^{n \times m}_+,
    v'_{i} \in R^m_+$,
    \begin{align}\label{def:dsic}\tag{\textrm{DSIC}}
      v_{i} \cdot x_{i}(\v) - p_{i}(\v)
        \geq v_{i} \cdot x_{i}(v'_{i}, \v_{-i})
                - p_{i}(v'_{i}, \v_{-i});
    \end{align}
    Ex-post Individually Rational (epIR), if $\forall \v \in \R^{n \times m}_+$,
    \begin{align}\label{def:epir}\tag{\textrm{epIR}}
      v_{i} \cdot x_{i}(\v) - p_{i}(\v) \geq 0.
    \end{align}

    Similarly, we have the following linear program for optimal
    dominant-strategy direct mechanisms:
    \begin{align}
      \max \quad & \sum_{\v} \mu(\v) \sum_i p_{i}(\v)
                   \label{lp:ds}\tag{\textrm{DSLP}}  \\
      \st  \quad & v_{i} \cdot x_{i}(\v) - p_{i}(\v)
                   \geq v_{i} \cdot x_{i}(v'_{i}, \v_{-i})
                        - p_{i}(v'_{i}, \v_{-i}),
                   ~\forall i \in [n],~\v \in \V,~v'_{i} \in \V_{i}
                   \nonumber  \\
                 & v_{i} \cdot x_{i}(\v) - p_{i}(\v) \geq 0,
                   ~\forall i \in [n],~v_{i} \in \V_{i}
                   \nonumber  \\
                 & \sum_i x\scr{i}{j}(\v) \leq 1,
                   ~\forall j \in [m],~\v \in \V \nonumber  \\
                 & x\scr{i}{j}(\v), p_{i}(\v) \geq 0,
                   ~\forall i \in [n],~j \in [m],~\v \in \V \nonumber
    \end{align}

    Again, any feasible solution to this linear program is only a limited
    dominant-strategy implementation, while the following lemma (first by
    \cite{DBLP:conf/stoc/DobzinskiFK11}) provides an extension method similar to
    the Bayesian case.

    \begin{lemma}[Dominant-Strategy Extension]
      Given any feasible solution $(\x, \p)$ to \ref{lp:ds}, the extended direct
      mechanism $(\x', \p')$ defined as follows satisfies \ref{def:dsic} and
      \ref{def:epir}.
      \begin{align*}
        &~\x'(\v) = \x(\v'),~ \p'(\v) = \p(\v')  \\
        \text{where}~\forall i \in [n],~&~\text{if}~\v \in \V,~\v' = \v  \\
        &~\text{else if}~\v_{-i} \in \V_{-i},~v'_i = \argmax_{v_i \in \V_i}
            v'_i \cdot x_i(v_i, \v_{-i}) - p_i(v_i, \v_{-i})  \\
        &~\text{otherwise},~v'_i = \Zero.
      \end{align*}
    \end{lemma}

    Therefore, from now on, we will only focus on the value space $\V$.

  \subsection{Separate Selling}

    By {\em separate-selling}, we mean to sell each of the item independently
    via the Myerson's auction. We use $\SRev$ to denote the revenue of
    separate-selling.

\section{Dual Programs and Virtual Values}\label{sec:dual}

  Now we write down the corresponding dual programs. In particular, we will omit
  the ``for-all'' quantifiers on the free variables in the rest of the paper.

  \subsection{Dominant-Strategy Implementation}
    \subsubsection{Duality and Complementary Slackness}

    First for dominant-strategy implementation, let $\zeta_{i}(v_{i},
    v'_{i}; \v_{-i})$, $\eta_{i}(\v)$, and $\xi\sci{j}(\v)$ be the
    corresponding multipliers of the constraints. By reorganizing \ref{lp:ds}
    into the standard form, we obtain,

    \paragraph{Primal Dominant-Strategy}
    \begin{align*}
                  & x\scr{i}{j}(\v), p_{i}(\v) \geq 0 & \textrm{variables}  \\
       \max \quad & \sum_{\v} \mu(\v) \sum_i p_{i}(\v) & \textrm{objective}  \\
       \st  \quad & - v_{i} \cdot x_{i}(\v) + p_{i}(\v)
                    + v_{i} \cdot x_{i}(v'_{i}, \v_{-i})
                         - p_{i}(v'_{i}, \v_{-i}) \leq 0
                    & \zeta_{i}(v_{i}, v'_{i}; \v_{-i})  \\
                  & - v_{i} \cdot x_{i}(\v) + p_{i}(\v) \leq 0
                    & \eta_{i}(\v)  \\
                  & \sum_i x\scr{i}{j}(\v) \leq 1 & \xi\sci{j}(\v)
    \end{align*}

    Hence the dual program is
    \paragraph{Dual Dominant-Strategy}
    \begin{equation}\label{dlp:ds}\tag{\textrm{Dual DSLP}}
      \begin{aligned}
        & \zeta_{i}(v_{i}, v'_{i}; \v_{-i}),
          \eta_{i}(\v),\xi\sci{j}(\v) \geq 0 & \textrm{variables}  \\
        \min \quad & \sum_{\v} \sum_j \xi\sci{j}(\v) & \textrm{objective}  \\
        \st  \quad & \xi\sci{j}(\v) - \left(\eta_{i}(\v)v\scr{i}{j}
            + \sum_{v'_{i}}\left(
              \zeta_{i}(v_{i}, v'_{i}; \v_{-i}) v\scr{i}{j}
              - \zeta_{i}(v'_{i}, v_{i}; \v_{-i})
                {v'}\scr{i}{j}
          \right)\right) \geq 0 & x\scr{i}{j}(\v)  \\
        & \eta_{i}(\v) + \sum_{v'_{i}}
            \left(\zeta_{i}(v_{i}, v'_{i}; \v_{-i})
                  - \zeta_{i}(v'_{i}, v_{i}; \v_{-i})\right)
          \geq \mu(\v) & p_{i}(\v)
      \end{aligned}
    \end{equation}

    We then use $\P$ to denote the polytope of all feasible solutions of the
    primal linear program, and similarly $\D$ for the polytope of the dual
    linear program. For ease of notation, we use the following abbreviations:
    \begin{align*}
      \textrm{For primal:} \quad &
          \Delta u_i(v_i, v'_i; \v_{-i})
            = v_i \cdot x_i(v'_i, \v_{-i}) - p_i(v'_i, \v_{-i}) -
                (v_i \cdot x_i(\v) + p_i(\v))  \\
        & u_i(\v) = v_i \cdot x_i(\v) + p_i(\v)  \\
        & s\sci{j}(\v) = \sum_i x\scr{i}{j}(\v)  \\
      \textrm{For dual:} \quad &
          \phi\scr{i}{j}(\v) =
            \eta_{i}(\v)v\scr{i}{j} + \sum_{v'_{i}}\left(
              \zeta_{i}(v_{i}, v'_{i}; \v_{-i}) v\scr{i}{j}
                - \zeta_{i}(v'_{i}, v_{i}; \v_{-i}){v'}\scr{i}{j}\right)  \\
        & \psi_i(\v) = \eta_{i}(\v) + \sum_{v'_{i}}
              \left(\zeta_{i}(v_{i}, v'_{i}; \v_{-i})
                    - \zeta_{i}(v'_{i}, v_{i}; \v_{-i})\right)
    \end{align*}

    Then the primal and dual look like
    \begin{align*}
      \begin{aligned}
        \textbf{Primal} \quad &
          x\scr{i}{j}(\v), p_{i}(\v) \geq 0 & \textrm{vars}  \\
        \max \quad &
          \sum_{\v} \mu(\v) \sum_i p_{i}(\v) & \textrm{obj}  \\
        \st  \quad & \Delta u_i(v_i, v'_i; \v_{-i}) \leq 0
                     & \zeta_{i}(v_{i}, v'_{i}; \v_{-i})  \\
                   & - u_i(\v) \leq 0  & \eta_{i}(\v)  \\
                   & s\sci{j}(\v) \leq 1 & \xi\sci{j}(\v)
      \end{aligned}
      \qquad \quad
      \begin{aligned}
        \textbf{Dual} \quad & \zeta_{i}(v_{i}, v'_{i}; \v_{-i}),
          \eta_{i}(\v),\xi\sci{j}(\v) \geq 0 & \textrm{vars}  \\
        \min \quad & \sum_{\v} \sum_j \xi\sci{j}(\v) & \textrm{obj}  \\
        \st  \quad &
          \xi\sci{j}(\v) - \phi\scr{i}{j}(\v) \geq 0 & x\scr{i}{j}(\v)  \\
                   & \psi_i(\v) \geq \mu(\v) & p_{i}(\v)  \\
                   & &
      \end{aligned}
    \end{align*}

    Clearly, both $\P$ and $\D$ are always nonempty, and $\P$ is bounded. Now,
    suppose $\pi^* = \langle{x^*}\scr{i}{j}(\v), p^*_{i}(\v)\rangle \in \P$ is
    an optimal solution of the primal, and $\delta^* = \langle\zeta^*_i(v_i,
    v'_i; \v_{-i}), \eta^*_i(\v),{\xi^*}\sci{j}(\v)\rangle \in \D$ is an optimal
    solution of the dual. By strong duality theorem, we know that $\obj(\pi^*) =
    \obj(\delta^*)$, which is the optimal revenue of any dominant-strategy
    implementation, denoted by $\DRev$:
    \begin{align*}
      \DRev = \obj(\pi^*) = \sum_{\v} \mu(\v) \sum_i p^*_i(\v)
            = \obj(\delta^*) = \sum_{\v} \sum_j {\xi^*}\sci{j}(\v).
    \end{align*}

    Finally, we add slack variables to both primal and dual:
    \begin{align*}
      \textbf{Primal} \quad &
        x\scr{i}{j}(\v), p_{i}(\v) \geq 0 & \textrm{variables}  \\
        & a_i(v_i, v'_i; \v_{-i}), b_i(\v), c\sci{j}(\v) \geq 0
        & \textrm{slack variables}  \\
      \max \quad &
        \sum_{\v} \mu(\v) \sum_i p_{i}(\v) & \textrm{objective}  \\
      \st  \quad & \Delta u_i(v_i, v'_i; \v_{-i})
                      + a_i(v_i, v'_i; \v_{-i}) = 0
                   & \zeta_{i}(v_{i}, v'_{i}; \v_{-i})  \\
                 & - u_i(\v) + b_i(\v) = 0  & \eta_{i}(\v)  \\
                 & s\sci{j}(\v) + c\sci{j}(\v) = 1 & \xi\sci{j}(\v)  \\
      & &  \\
      \textbf{Dual} \quad & \zeta_{i}(v_{i}, v'_{i}; \v_{-i}),
        \eta_{i}(\v),\xi\sci{j}(\v) \geq 0 & \textrm{variables}  \\
        & \alpha\scr{i}{j}(\v), \beta_i(\v) \geq 0 & \textrm{slack variables}
          \\
      \min \quad & \sum_{\v} \sum_j \xi\sci{j}(\v) & \textrm{objective}  \\
      \st  \quad &
        \xi\sci{j}(\v) - \phi\scr{i}{j}(\v) - \alpha\scr{i}{j}(\v) = 0
                 & x\scr{i}{j}(\v)  \\
                 & \psi_i(\v) - \beta_i(\v) = \mu(\v) & p_{i}(\v)
    \end{align*}

    In what follows, we will abuse the notation $\P$ and $\D$ as the feasible
    polytopes for both normal variables and slack variables for primal and dual,
    respectively.

    \paragraph{Complementary Slackness}
    For any feasible primal solution $\pi \in \P$ and dual solution $\delta \in
    \D$, they are the optimal solution for primal and dual if and only if:
    \begin{gather*}
      a_i(v_i, v'_i; \v_{-i}) \zeta_{i}(v_{i}, v'_{i}; \v_{-i}) = 0,~
      b_i(\v) \eta_{i}(\v) = 0,~c\sci{j}(\v) \xi\sci{j}(\v) = 0  \\
      \alpha\scr{i}{j}(\v) x\scr{i}{j}(\v) = 0,~\beta_i(\v) p_{i}(\v) = 0.
    \end{gather*}

    In particular, the gap between primal and dual objectives equals the sum of
    all the products.
    \begin{align}\label{eq:ds-slack}\tag{\textrm{DSCS}}
      \obj(\delta) - \obj(\pi) =&
        \sum_{i,\v,v'_i \neq v_i}
          a_i(v_i, v'_i; \v_{-i}) \zeta_{i}(v_{i}, v'_{i}; \v_{-i})
          + \sum_{i,\v} b_i(\v) \eta_{i}(\v)  \\
        & + \sum_{j,\v} c\sci{j}(\v) \xi\sci{j}(\v)
          + \sum_{i,j,\v} \alpha\scr{i}{j}(\v) x\scr{i}{j}(\v)
          + \sum_{i,\v} \beta_i(\v) p_{i}(\v). \nonumber
    \end{align}

    We then focus on interpreting the complementary slackness conditions.

    \subsubsection{Virtual Values}

    For any fixed optimal solutions $\pi^* \in \P$ and $\delta^* \in \D$, since
    ${\alpha^*}\scr{i}{j}(\v) {x^*}\scr{i}{j}(\v) = 0$, we conclude that:
    \begin{align*}
      {x^*}\scr{i}{j}(\v) > 0 \Infer 0 = {\alpha^*}\scr{i}{j}(\v)
        = {\xi^*}\sci{j}(\v) - {\phi^*}\scr{i}{j}(\v).
    \end{align*}

    In particular, it also implies that item $j$ is only allocated to the
    buyer(s) $i$ with maximum ${\phi^*}\scr{i}{j}(\v)$:
    \begin{align*}
      \forall i' \in [n],~{\phi^*}\scr{i}{j}(\v) = {\xi^*}\sci{j}(\v)
        \geq {\phi^*}\scr{i'}{j}(\v),
    \end{align*}
    where the inequality is the first constraint of the dual program
    ($\xi\sci{j}(\v) - \phi\scr{i}{j}(\v) \geq 0$).

    Moreover, by ${c^*}\sci{j}(\v) {\xi^*}\sci{j}(\v) = 0$, we obtain:
    \begin{align*}
      1 - \sum_i {x^*}\scr{i}{j} = {c^*}\sci{j}(\v) > 0 \Infer
        0 = {\xi^*}\sci{j}(\v) \geq {\phi^*}\scr{i'}{j}(\v),~\forall i' \in [n],
    \end{align*}
    which implies that item $j$ is not fully allocated only if for all buyer
    $i'$, ${\phi^*}\scr{i'}{j}(\v)$ is not strictly positive.

    \begin{definition}[Expected Virtual Values]\label{def:evirtual}
      For any optimal dual solution $\delta^*$, ${\phi^*}\scr{i}{j}(\v)$ defines
      the {\em expected virtual values} in the sense that any optimal auction
      must maximize the expected virtual welfare.
    \end{definition}

    Note that the expected virtual values are different from the virtual values
    commonly used in revenue maximization literatures. For example, virtual
    values are not always well-defined for some extreme cases (such as
    discrete/unbounded distributions and distributions with point masses), while
    the expected virtual values are always explicitly defined by an dual optimal
    solution. Later in \autoref{thm:virtual}, we will formally define the
    virtual values in our setting as analog to common virtual values (e.g.,
    Myerson's virtual value).

    \begin{theorem}[Virtual Values]
      \label{thm:virtual}
      For any optimal dual solution $\delta^*$ satisfying certain regularization
      conditions (defined later), there exists corresponding virtual value
      functions $\varphi\scr{i}{j}: \V \rightarrow \R \cup \{-\infty\}$ such
      that
      \begin{align*}
        \varphi\scr{i}{j}(\v)\mu(\v) = {\phi^*}\scr{i}{j}(\v).
      \end{align*}

      In particular,
      \begin{align*}
        \varphi\scr{i}{j}(\v) = \left\{\begin{array}{ll}
          {\phi^*}\scr{i}{j}(\v) / \mu(\v), & \text{if}~\mu(\v) > 0;  \\
          \text{some real number in }\R, & \text{if}~\mu_{-i}(\v_{-i}) = 0;  \\
          -\infty, & \text{if}~v_i = \Zero,~\mu_i(\Zero) = 0,~
                     \mu_{-i}(\v_{-i}) > 0.
        \end{array}\right.
      \end{align*}

      Moreover, any optimal auction (optimal primal solution) $\pi^*$ must be a
      virtual welfare maximizer:
      \begin{enumerate}
        \itemsep0em
        \item Each item is only allocated to the buyer(s) with the highest
              and non-negative virtual value on this item;
        \item The highest virtual value for any unallocated (or partially
              allocated) item must be non-positive (for partially allocated
              items, the highest virtual value must be zero).
      \end{enumerate}
    \end{theorem}

    The basic idea is to simply define $\varphi\scr{i}{j}(\v) =
    {\phi^*}\scr{i}{j}(\v) / \mu(\v)$, while it only works when $\mu(\v) > 0$.
    For those $\v \in \V$ such that $\mu(\v) = 0$, the virtual value
    $\varphi\scr{i}{j}(\v)$ can be defined only if ${\phi^*}\scr{i}{j}(\v) = 0$
    as well.

    Hence the first regularization condition is:
    \begin{align}\label{eq:regular-virtual}
      \forall \v \in \V,~i \in [n],~\mu_{-i}(\v_{-i}) = 0
        \Infer {\phi^*}\scr{i}{j}(\v) = 0.
    \end{align}

    Besides, we will require two more regularization conditions to simplify the
    discussion in upcoming sections, which are:
    \begin{gather}
      \forall \v \in \V,~i \in [n],~v_i \neq \Zero,~\eta^*_i(\v) = 0~\text{and}~
      \eta^*_i(\Zero, \v_{-i}) = \mu_{-i}(\v_{-i}),
        \label{eq:regular-source}  \\
      \forall \v \in \V,~i \in [n],~\beta^*_i(\v) = \psi^*_i(\v) - \mu(\v) = 0.
        \label{eq:regular-trans}
    \end{gather}

    By our previous interpretations on some of the complementary slackness
    conditions, we remain to prove the following lemma:
    \begin{lemma}[Regular Dual OPT]\label{lem:regular-exist}
      There always exists an optimal dual solution $\delta^*$ satisfying the
      regularization condition \eqref{eq:regular-virtual},
      \eqref{eq:regular-source}, and \eqref{eq:regular-trans}.\footnote{We note
      that some of the optimal dual solution $\delta^*$ may not satisfy these
      regularization, but there always exist the ones satisfy them.}
    \end{lemma}

    \begin{proofof}{\autoref{thm:virtual}}
      Directly implied by \autoref{lem:regular-exist}.
    \end{proofof}

    Furthermore, by \autoref{lem:regular-exist}, we can reformulate
    ${\phi^*}\scr{i}{j}(\v)$ as follows,
    \begin{align}
      &
      {\phi^*}\scr{i}{j}(\v)
        = \psi^*_i(\v)v\scr{i}{j} + \sum_{v'_{i}}
          \zeta^*_{i}(v'_{i}, v_{i}; \v_{-i})(v\scr{i}{j} - {v'}\scr{i}{j})
        = \mu(\v)v\scr{i}{j} + \sum_{v'_{i}}
          \zeta^*_{i}(v'_{i}, v_{i}; \v_{-i})(v\scr{i}{j} - {v'}\scr{i}{j})
          \nonumber  \\
      \Infer~&
      \varphi\scr{i}{j}(\v) = v\scr{i}{j} + \sum_{v'_{i}}
        \zeta^*_{i}(v'_{i}, v_{i}; \v_{-i})(v\scr{i}{j} - {v'}\scr{i}{j})
        / \mu(\v).
      \label{eq:ds-virtual}\tag{\textrm{DSVV}}
    \end{align}

  \subsection{Bayesian Implementation}
    \subsubsection{Duality and Complementary Slackness}

    Now, for Bayesian implementation, let $\bar \zeta_{i}(v_{i}, v'_{i})$,
    $\bar \eta_{i}(v_i)$, and $\bar \xi\sci{j}(\v)$ be the
    corresponding multipliers of the constraints. By reorganizing \ref{lp:bayes}
    into the standard form, we obtain,
    \paragraph{Primal Bayesian}
    \begin{align*}
                  & \bar x\scr{i}{j}(\v), \bar p_{i}(\v) \geq 0
                  & \textrm{variables}  \\
       \max \quad & \sum_{\v} \mu(\v) \sum_i \bar p_{i}(\v)
                  & \textrm{objective}  \\
       \st  \quad & \sum_{\v_{-i}} \mu_{-i}(\v_{-i})
                      (- v_{i} \cdot \bar x_{i}(\v) + \bar p_{i}(\v)
                        + v_{i} \cdot \bar x_{i}(v'_{i}, \v_{-i})
                          - \bar p_{i}(v'_{i}, \v_{-i})) \leq 0
                    & \bar \zeta_{i}(v_{i}, v'_{i})  \\
                  & \sum_{\v_{-i}} \mu_{-i}(\v_{-i})
                      (-v_{i} \cdot x_{i}(\v) + \bar p_{i}(\v)) \leq 0
                    & \bar \eta_{i}(v_i)  \\
                  & \sum_i \bar x\scr{i}{j}(\v) \leq 1 & \bar \xi\sci{j}(\v)
    \end{align*}

    Hence the dual program is
    \paragraph{Dual Bayesian}
    \begin{equation}\label{dlp:bayes}\tag{\textrm{Dual BLP}}
      \begin{aligned}
                    & \bar \zeta_{i}(v_{i}, v'_{i}),
                      \bar \eta_{i}(v_i), \bar \xi\sci{j}(\v) \geq 0
                    & \textrm{variables}  \\
         \min \quad & \sum_{\v} \sum_j \bar \xi\sci{j}(\v) & \textrm{objective}
                      \\
         \st  \quad & \bar \xi\sci{j}(\v)
                        - \mu_{-i}(\v_{-i})\left(\bar \eta_{i}(v_i)v\scr{i}{j}
                        + \sum_{v'_{i}}\left(
                          \bar \zeta_{i}(v_{i}, v'_{i}) v\scr{i}{j}
                          - \bar \zeta_{i}(v'_{i}, v_{i}) {v'}\scr{i}{j}
                      \right)\right) \geq 0 & \bar x\scr{i}{j}(\v)  \\
                    & \mu_{-i}(\v_{-i})\left(\bar \eta_{i}(v_i) + \sum_{v'_{i}}
                        \left(\bar \zeta_{i}(v_{i}, v'_{i})
                              - \bar \zeta_{i}(v'_{i}, v_{i})\right)\right)
                      \geq \mu(\v) & \bar p_{i}(\v)
      \end{aligned}
    \end{equation}

    Similarly, we then use $\bar \P$ to denote the polytope of all feasible
    solutions of the Bayesian primal linear program, $\bar \D$ for the
    polytope of the Bayesian dual linear program, and the following
    abbreviations:
    \begin{align*}
      \textrm{For primal:} \quad &
          \Delta \bar u_i(v_i, v'_i)
            = \sum_{\v_{-i}} \mu_{-i}(\v_{-i})
                \left(v_i \cdot \bar x_i(v'_i, \v_{-i}) - \bar p_i(v'_i, \v_{-i}) -
                (v_i \cdot \bar x_i(\v) + \bar p_i(\v))\right)  \\
        & \bar u_i(v_i) = \sum_{\v_{-i}} \mu_{-i}(\v_{-i})
                             (v_i \cdot \bar x_i(\v) + \bar p_i(\v))  \\
        & \bar s\sci{j}(\v) = \sum_i \bar x\scr{i}{j}(\v)  \\
      \textrm{For dual:} \quad &
          \bar \phi\scr{i}{j}(v_i) =
            \bar \eta_{i}(v_i)v\scr{i}{j} + \sum_{v'_{i}}\left(
              \bar \zeta_{i}(v_{i}, v'_{i}) v\scr{i}{j}
                - \bar \zeta_{i}(v'_{i}, v_{i}){v'}\scr{i}{j}\right)  \\
        & \bar \psi_i(v_i) = \bar \eta_{i}(v_i) + \sum_{v'_{i}}
              \left(\bar \zeta_{i}(v_{i}, v'_{i})
                    - \bar \zeta_{i}(v'_{i}, v_{i})\right)
    \end{align*}
    %
    %
    Clearly, both $\bar \P$ and $\bar \D$ are always nonempty, and $\bar \P$ is
    bounded. Now, suppose $\bar \pi^* = \langle{\bar {x^*}}\scr{i}{j}(\v),
    \bar p^*_{i}(\v)\rangle \in \bar \P$ is an optimal solution of the primal,
    and $\bar \delta^* = \langle\bar \zeta^*_i(v_i, v'_i), \bar \eta^*_i(v_i),
    \bar {\xi^*}\sci{j}(\v)\rangle \in \bar \D$ is an optimal solution of the
    dual. By strong duality theorem, we know that $\obj(\bar \pi^*) =
    \obj(\bar \delta^*)$, which is the optimal revenue of any Bayesian
    implementation, denoted by $\BRev$:
    \begin{align*}
      \BRev = \obj(\bar \pi^*) = \sum_{\v} \mu(\v) \sum_i \bar p^*_i(\v)
            = \obj(\bar \delta^*) = \sum_{\v} \sum_j \bar {\xi^*}\sci{j}(\v).
    \end{align*}

    Again, we add slack variables to both primal and dual:
    \begin{align*}
      \textbf{Primal} \quad &
        \bar x\scr{i}{j}(\v), \bar p_{i}(\v) \geq 0 & \textrm{variables}  \\
        & \bar a_i(v_i, v'_i), \bar b_i(v_i), \bar c\sci{j}(\v) \geq 0
        & \textrm{slack variables}  \\
      \max \quad &
        \sum_{\v} \mu(\v) \sum_i \bar p_{i}(\v) & \textrm{objective}  \\
      \st  \quad & \Delta \bar u_i(v_i, v'_i) + \bar a_i(v_i, v'_i) = 0
                   & \bar \zeta_{i}(v_{i}, v'_{i})  \\
                 & - \bar u_i(v_i) + \bar b_i(v_i) = 0
                   & \bar \eta_{i}(v_i)  \\
                 & \bar s\sci{j}(\v) + \bar c\sci{j}(\v) = 1
                   & \bar \xi\sci{j}(\v)  \\
      & &  \\
      \textbf{Dual} \quad & \bar \zeta_{i}(v_{i}, v'_{i}),
        \bar \eta_{i}(v_i), \bar \xi\sci{j}(\v) \geq 0 & \textrm{variables}  \\
        & \bar \alpha\scr{i}{j}(\v), \bar \beta_i(\v) \geq 0
          & \textrm{slack variables}  \\
      \min \quad & \sum_{\v} \sum_j \bar \xi\sci{j}(\v) & \textrm{objective}  \\
      \st  \quad &
        \bar \xi\sci{j}(\v) - \mu_{-i}(\v_{-i})\bar \phi\scr{i}{j}(v_i)
          - \bar \alpha\scr{i}{j}(\v) = 0
                 & \bar x\scr{i}{j}(\v)  \\
                 & \mu_{-i}(\v_{-i})\bar \psi_i(v_i) - \bar \beta_i(\v) = \mu(\v)
                   & \bar p_{i}(\v)
    \end{align*}

    As we did for dominant-strategy implementation, we will abuse the notation
    $\bar \P$ and $\bar \D$ as the feasible polytopes for both normal variables
    and slack variables of primal and dual, respectively.

    \paragraph{Complementary Slackness}
    For any feasible primal solution $\bar \pi \in \bar \P$ and dual solution
    $\bar \delta \in \bar \D$, they are the optimal solution for primal and dual
    if and only if:
    \begin{gather*}
      \bar a_i(v_i, v'_i) \bar \zeta_{i}(v_{i}, v'_{i}) = 0,~
      \bar b_i(v_i) \bar \eta_{i}(v_i) = 0,~
      \bar c\sci{j}(\v) \bar \xi\sci{j}(\v) = 0  \\
      \bar \alpha\scr{i}{j}(\v) \bar x\scr{i}{j}(\v) = 0,~
      \bar \beta_i(\v) \bar p_{i}(\v) = 0.
    \end{gather*}

    In particular, the gap between primal and dual objectives equals the sum of
    all the products.
    \begin{align}\label{eq:bayes-slack}\tag{\textrm{BCS}}
      \obj(\bar \delta) - \obj(\bar \pi) =&
        \sum_{i,v'_i \neq v_i}
          \bar a_i(v_i, v'_i) \bar \zeta_{i}(v_{i}, v'_{i})
          + \sum_{i,v_i} \bar b_i(v_i) \bar \eta_{i}(v_i)  \\
        & + \sum_{j,\v} \bar c\sci{j}(\v) \bar \xi\sci{j}(\v)
          + \sum_{i,j,\v} \bar \alpha\scr{i}{j}(\v) \bar x\scr{i}{j}(\v)
          + \sum_{i,\v} \bar \beta_i(\v) \bar p_{i}(\v). \nonumber
    \end{align}

    \subsubsection{(Bayesian) Virtual Values}

    We then repeat the interpreting of the complementary slackness conditions as
    we did for the dominant-strategy implementation. In particular, we can
    conclude that $\mu_{-i}(\v_{-i}) \bar {\phi^*}\scr{i}{j}(\v)$ is the
    expected virtual value in Bayesian setting:

    \begin{definition}[Expected (Bayesian) Virtual Values]\label{def:ebvirtual}
      For any optimal dual solution $\bar \delta^*$,
      $\mu_{-i}(\v_{-i}) \bar {\phi^*}\scr{i}{j}(\v)$ defines the {\em expected
      (Bayesian) virtual values} in the sense that any optimal auction must
      maximize the expected virtual welfare.
    \end{definition}
    %
    %

    Similarly, we can define virtual values for Bayesian implementations.

    \begin{theorem}[(Baysian) Virtual Values]
      \label{thm:bayes-virtual}
      For any optimal dual solution $\bar \delta^*$ satisfying certain
      regularization conditions (defined later), there exists corresponding
      virtual value functions $\bar \varphi\scr{i}{j}: \V_i \rightarrow \R \cup
      \{-\infty\}$ such that
      \begin{align*}
        \bar \varphi\scr{i}{j}(v_i)\mu(\v)
          = \bar {\phi^*}\scr{i}{j}(v_i) \mu_{-i}(\v_{-i}).
      \end{align*}

      In particular,
      \begin{align*}
        \bar \varphi\scr{i}{j}(v_i) = \left\{\begin{array}{ll}
          \bar {\phi^*}\scr{i}{j}(v_i) / \mu_i(v_i), & \text{if}~\mu(\v) > 0;  \\
          \text{some real number in }\R, & \text{if}~\mu_{-i}(\v_{-i}) = 0;  \\
          -\infty, & \text{if}~v_i = \Zero,~\mu_i(\Zero) = 0,~
                     \mu_{-i}(\v_{-i}) > 0.
        \end{array}\right.
      \end{align*}

      Moreover, any optimal auction (optimal primal solution) $\bar \pi^*$ must
      be a virtual welfare maximizer:
      \begin{enumerate}
        \itemsep0em
        \item Each item is only allocated to the buyer(s) with the highest
              and non-negative virtual value on this item;
        \item The highest virtual value for any unallocated (or partially
              allocated) item must be non-positive (for partially allocated
              items, the highest virtual value must be zero).
      \end{enumerate}
    \end{theorem}

    The corresponding regularization conditions are as follows:
    \begin{gather}
      \forall \v \in \V,~i \in [n],~\mu_{-i}(\v_{-i}) = 0
        \Infer \bar {\phi^*}\scr{i}{j}(v_i) = 0
        \label{eq:bayes-regular-virtual}  \\
      \forall \v \in \V,~i \in [n],~v_i \neq \Zero,~
        \bar \eta^*_i(v_i) = 0~\text{and}~
      \bar \eta^*_i(0) = 1,
        \label{eq:bayes-regular-source}  \\
      \forall \v \in \V,~i \in [n],~
        \bar \beta^*_i(\v) = \mu_{-i}(\v_{-i})\bar \psi^*_i(v_i) - \mu(\v) = 0.
        \label{eq:bayes-regular-trans}
    \end{gather}
    %
    %
    %
    %

    By our previous interpretations on some of the complementary slackness
    conditions, we remain to prove the following lemma:
    \begin{lemma}[Regular Dual OPT]\label{lem:bayes-regular-exist}
      There always exists an optimal dual solution $\delta^*$ satisfying the
      regularization condition \eqref{eq:bayes-regular-virtual},
      \eqref{eq:bayes-regular-source}, and \eqref{eq:bayes-regular-trans}.\footnote{We note
      that some of the optimal dual solution $\delta^*$ may not satisfy these
      regularization, but there always exist the ones satisfy them.}
    \end{lemma}

    \begin{proofof}{\autoref{thm:bayes-virtual}}
      Directly implied by \autoref{lem:bayes-regular-exist}.
    \end{proofof}

    Furthermore, by \autoref{lem:bayes-regular-exist}, we can reformulate
    $\bar {\phi^*}\scr{i}{j}(v_i)$ as follows,
    \begin{align}
      &
      \bar {\phi^*}\scr{i}{j}(v_i)
        = \bar \psi^*_i(v_i)v\scr{i}{j} + \sum_{v'_{i}}
          \bar \zeta^*_{i}(v'_{i}, v_{i})(v\scr{i}{j} - {v'}\scr{i}{j})
        = \mu_i(v_i)v\scr{i}{j} + \sum_{v'_{i}}
          \bar \zeta^*_{i}(v'_{i}, v_{i})(v\scr{i}{j} - {v'}\scr{i}{j})
          \nonumber  \\
      \Infer~&
      \bar \varphi\scr{i}{j}(v_i) = v\scr{i}{j} + \sum_{v'_{i}}
        \bar \zeta^*_{i}(v'_{i}, v_{i})(v\scr{i}{j} - {v'}\scr{i}{j})
        / \mu_i(v_i).
      \label{eq:bayes-virtual}\tag{\textrm{BVV}}
    \end{align}

\section{Characterization}\label{sec:charac}

  In this section, we present the sufficient and necessary characterization of
  BIC $=$ DSIC.

  \subsection{A sufficient and necessary condition of BIC $=$ DSIC}

    In previous sections, we defined two types of virtual values, i.e.,
    dominant-strategy virtual values \eqref{eq:ds-virtual} and Bayesian virtual
    values \eqref{eq:bayes-virtual}. In particular, the Bayesian virtual values
    for buyer $i$, $\bar \varphi\scr{i}{j}(v_i)$, are independent of the values
    of other buyers by the construction, while the dominant-strategy virtual
    values for buyer $i$, $\varphi\scr{i}{j}(\v)$, depend on the values of other
    buyers as well.

    We say the (regular) dominant-strategy virtual values are {\em
    agent-independent}, if for each buyer $i$, her virtual values (and related
    dual variables,  $\eta_i(\v)$ and $\zeta_i(v_i, v'_i; \v_{-i})$) are
    independent of the values of other buyers:
    \begin{equation}\label{def:agent-inde}\tag{\textrm{AI}}
      \begin{gathered}
        \forall v_i \in \V_i, \v_{-i}, \v'_{-i}, \in \V_{-i}, \qquad
          \varphi_i(v_i, \v_{-i}) = \varphi_i(v_i, \v'_{-i}),  \\
          \eta_i(\v) \mu_{-i}(\v'_{-i})
            = \eta_i(v_i, \v'_{-i}) \mu_{-i}(\v_{-i}), \quad
          \zeta_i(v_i, v'_i; \v_{-i}) \mu_{-i}(\v'_{-i})
            = \zeta_i(v_i, v'_i; \v'_{-i}) \mu_{-i}(\v_{-i}).
      \end{gathered}
    \end{equation}

    Then our first main result is the following characterization:
    \begin{theorem}\label{thm:iff}
      BIC $=$ DSIC if and only if that there is an optimal DSIC auction that is
      induced by agent-independent virtual values.
    \end{theorem}

    \begin{proofof}{\autoref{thm:iff}}
      ``$\Infer$'':

      One key observation is that any solution of \ref{dlp:bayes} induces a
      solution of \ref{dlp:ds}. In particular, let $\bar \delta^* =
      \langle \bar \zeta^*_{i}(v'_{i}, v_{i}), \bar \eta^*_i(v_i),
      \bar {\xi^*}\sci{j}(\v) \rangle \in \bar \D$ be an optimal solution to
      \ref{dlp:bayes}. The following $\hat \delta$ constructed from $\bar
      \delta^*$ is a feasible solution to \ref{dlp:ds}:
      \begin{align*}
        \hat \delta &= \langle \hat \zeta_i(v_i, v'_i; \v_{-i}),
                          \hat \eta_i(\v), \hat \xi\sci{j}(\v)\rangle  \\
        \hat \zeta_i(v_i, v'_i; \v_{-i})
            &= \bar \zeta^*_{i}(v'_{i}, v_{i}) \mu_{-i}(\v_{-i})  \\
        \hat \eta_i(\v)
            &= \bar \eta^*_i(v_i) \mu_{-i}(\v_{-i})  \\
        \hat \xi\sci{j}(\v)
            &= \bar {\xi^*}\sci{j}(\v).
      \end{align*}

      We omit the verification of $\hat \delta \in \D$, which is directly
      implied by the definition of \ref{dlp:bayes} and \ref{dlp:ds} (as well as
      the fact that $\mu_{-i}(\v_{-i}) \geq 0$).

      In the meanwhile, note that the objective value of $\bar \delta^*$ in
      \ref{dlp:bayes} is the same as the objective value of $\hat \delta$ in
      \ref{dlp:ds}, we conclude that:
      \begin{align*}
        \obj(\bar \delta^*) = \obj(\hat \delta) \geq \obj(\delta^*),
      \end{align*}
      where $\delta^*$ is an optimal solution of \ref{dlp:ds} and the last
      inequality is from the optimality of $\delta^*$.

      On the other hand, by the hypothesis that BIC $=$ DSIC, i.e.,
      $\obj(\bar \delta^*) = \obj(\delta^*)$, the constructed solution $\hat
      \delta$ is in fact an optimal solution to \ref{dlp:ds}.

      Since $\bar \delta^*$ is an arbitrary optimal solution to \ref{dlp:bayes},
      we can further assume that it is regular
      (\autoref{lem:bayes-regular-exist}). The corresponding $\hat \delta$ then
      is also regular according to the definition in \autoref{lem:regular-exist}
      (we omit the verification here, which is straightforward by the
      definitions). Therefore, $\hat \delta$ defines the virtual values for an
      optimal DSIC auction \eqref{eq:ds-virtual}:
      \begin{align*}
        \hat \varphi\scr{i}{j}(\v)
          & = v\scr{i}{j} + \sum_{v'_{i}}
          \hat \zeta_{i}(v'_{i}, v_{i}; \v_{-i})(v\scr{i}{j} - {v'}\scr{i}{j})
          / \mu(\v)  \\
          & = v\scr{i}{j} + \sum_{v'_{i}}
          \bar \zeta^*_{i}(v'_{i}, v_{i})(v\scr{i}{j} - {v'}\scr{i}{j})
          / \mu_i(v_i).
      \end{align*}

      In particular, the virtual values are the same as \eqref{eq:bayes-virtual}
      and are agent-independent. ($\hat \eta_i(\v) / \mu_{-i}(\v_{-i})$ and
      $\hat \zeta_i(v_i, v'_i; \v_{-i}) / \mu_{-i}(\v_{-i})$ are also invariant
      in $\v_{-i}$.)
      \medskip

      ``$\Reduce$'':

      By the hypothesis that there exist agent-independent virtual values
      $\delta^* = \langle \zeta^*_{i}(v'_{i}, v_{i}; \v_{-i}), \eta^*_i(\v),
      {\xi^*}\sci{j}(\v) \rangle$ inducing an optimal DSIC auction, we can
      construct the following $\tld \delta$, which is a feasible solution to
      \ref{dlp:bayes}:
      \begin{align*}
        \tld \delta & = \langle \tld \zeta_{i}(v'_{i}, v_{i}; \v_{-i}),
          \tld \eta_i(\v), \tld \xi\sci{j}(\v) \rangle  \\
        \tld \zeta_{i}(v'_{i}, v_{i}) & =
          \zeta^*_{i}(v'_{i}, v_{i}; \v_{-i}) / \mu_{-i}(\v_{-i})  \\
        \tld \eta_i(v_i) & = \eta^*_i(\v) / \mu_{-i}(\v_{-i})  \\
        \tld \xi\sci{j}(\v) & = {\xi^*}\sci{j}(\v).
      \end{align*}

      Note that according to the definiton of agent-independence
      \eqref{def:agent-inde}, the construction of $\tld \delta$ is consistent
      for all $\v_{-i}$. In particular, if $\mu_{-i}(\v_{-i}) = 0$, for all
      $\v'_{-i} \in \V_{-i}$, by \eqref{def:agent-inde},
      \begin{align*}
        \zeta^*_{i}(v'_{i}, v_{i}; \v_{-i}) \mu_{-i}(\v'_{-i})
          = \zeta^*_{i}(v'_{i}, v_{i}; \v'_{-i}) \mu_{-i}(\v_{-i}) = 0.
      \end{align*}

      Then $\zeta^*_{i}(v'_{i}, v_{i}; \v_{-i})$ must be zero, as there exists
      $\v'_{-i}$ such that $\mu_i(\v'_{-i}) > 0$. Hence in such special cases,
      we can safely define $\tld \zeta_{i}(v'_{i}, v_{i}) = 0$ and similarly
      $\tld \eta_i(v_i) = 0$.

      In the meanwhile, such a construction ensures that (i) its objective value
      in \ref{lp:bayes} being the same as the objective value of $\delta^*$ in
      \ref{lp:ds}
      \begin{align*}
        \obj(\tld \delta) = \obj(\delta^*),
      \end{align*}
      and (ii) $\tld \delta$ is also a feasible solution to \ref{dlp:bayes}
      (we omit the further verification here, which is straightforward by the
      definitions)
      \begin{gather*}
        \tld \zeta_{i}(v'_{i}, v_{i}) \mu_{-i}(\v_{-i})
          = \zeta^*_{i}(v'_{i}, v_{i}; \v_{-i}) \qquad
        \tld \eta_i(v_i) \mu_{-i}(\v_{-i}) = \eta^*_i(\v).
      \end{gather*}

      Therefore, we conclude that DSIC $=$ BIC:
      \begin{align*}
        \obj(\tld \delta) \geq \obj(\bar \delta^*) \geq \obj(\delta^*)
          = \obj(\tld \delta),
      \end{align*}
      where $\bar \delta^*$ is any optimal solution to \ref{dlp:bayes} and the
      last inequality is due to the fact that BIC $\geq$ DSIC.
    \end{proofof}

  \section{The I.I.D. Setting}\label{sec:iid}

    In this section, we further show that if the value distributions are i.i.d.
    and $0$ is in the supports, then the previous characterization implies that
    $\DRev = \SRev$. In the meanwhile, since separate selling employees
    agent-independent virtual values, $\DRev = \SRev$ directly implies that
    $\BRev = \DRev = \SRev$. In other words, although $\BRev \geq \DRev \geq
    \SRev$ in general, any two of them being equal implies that all of them are
    equal:
    \begin{corollary}\label{coro:iff}
      For $n \geq 3$,
      \begin{align*}
        \BRev = \DRev \quad \text{or} \quad \DRev = \SRev \quad \text{or} \quad
        \SRev = \BRev \quad \Infer \quad \BRev = \DRev = \SRev.
      \end{align*}
    \end{corollary}

    In fact, we have the following theorem:

    \begin{theorem}\label{thm:agent-item}
      In the i.i.d. value setting with $n \geq 3$, if the optimal DSIC auction
      is induced by agent-independent virtual values, there exist
      item-independent (and agent-independent) virtual values inducing an
      optimal DSIC auction.
    \end{theorem}

    The agent-independent virtual values are called {\em item-independent}, if
    \begin{align*}
      \forall v_i, v'_i \in \V_i, \quad
      \varphi\scr{i}{j}(v_i, v\scr{i}{-j})
        = \varphi\scr{i}{j}(v_i, {v'}\scr{i}{-j}) \quad \text{or} \quad
      \varphi\scr{i}{j}(v_i, v\scr{i}{-j}),
      \varphi\scr{i}{j}(v_i, {v'}\scr{i}{-j}) \leq 0.
    \end{align*}

    In particular, \autoref{coro:iff} directly follows from
    \autoref{thm:agent-item}:
    \begin{itemize}
      \item if virtual value $\varphi\scr{i}{j}$ is restricted to depending on
            $v\scr{i}{j}$ only, separate selling via Myerson's auction would be
            the optimal (hence $\BRev = \DRev \Infer \BRev = \DRev = \SRev$);
      \item if $\DRev = \SRev$, then the optimal DSIC auction can be induced by
            agent- and item-independent virtual values, implying DSIC $=$ BIC
            (hence $\DRev = \SRev \Infer \BRev = \DRev = \SRev$);
      \item if $\SRev = \BRev$, then by $\BRev \geq \DRev \geq \SRev$, all of
            them must be equal (hence $\SRev = \BRev \Infer \BRev = \DRev =
            \SRev$).
    \end{itemize}

    We then move to the proof of \autoref{thm:agent-item}, which relies on the
    following lemma:
    \begin{lemma}[Upper Bounded Virtual Values]\label{lem:ubvv}
      $\varphi\scr{i}{j} \leq v\scr{i}{j}$.
    \end{lemma}

    \begin{proofof}{\autoref{thm:agent-item}}
      We prove by contradiction. Assume that the virtual values are not
      item-independent. Note that the valuations are i.i.d., hence, without loss
      of generality, we assume that the virtual values for the agents are the
      same and the allocations are symmetric. Hence we also omit the subscripts
      of virtual values throught the proof.

      In particular, let $\bar v^j$ denote the maximum value of the $j$-th item
      in the support and $v\scr{0}{j}$ denote the value profile with all maximum
      value except that the value of the $j$-th item being $0$:
      \begin{align*}
        v\scr{0}{j}
          = (\bar v^1, \ldots, \bar v^{j-1}, 0, \bar v^{j+1}, \ldots, \bar v^m).
      \end{align*}

      Consider $v_i,v'_i$ and $j$ such that
      $\varphi^j(v\scr{i}{j}, v\scr{i}{-j})$ and
      $\varphi^j(v\scr{i}{j}, {v'}\scr{i}{-j})$ are different, i.e.,
      \begin{align}\label{eq:assump}
        \varphi^j(v\scr{i}{j}, v\scr{i}{-j})
          < \varphi^j(v\scr{i}{j}, {v'}\scr{i}{-j}) \quad \text{and} \quad
          \varphi^j(v\scr{i}{j}, {v'}\scr{i}{-j}) > 0.
      \end{align}

      Let $v = v_i$ and $v^{(-j)} = (v\scr{i}{j}, {v'}\scr{i}{-j})$. Pick an
      arbitrary buyer $i' \neq i$ and fix her values being $v\scr{0}{j}$. Then
      fix the values of all the remaining buyers (except for $i$ and $i'$)
      being $v$, i.e.,
      \begin{align*}
        v_{i''} = v, \forall i'' \neq i, i'.
      \end{align*}

      Note that the virtual values of buyer $i'$ are already determined.
      According to the definition of agent-independent virtual values:
      \begin{align*}
        \varphi\scr{i}{j}(v_i)
          = v\scr{i}{j} + \sum_{v'_i}
                \zeta^*_i(v'_i, v_i)(v\scr{i}{j} - {v'}\scr{i}{j}) / \mu_i(v_i),
      \end{align*}
      we have that $\varphi^j(v\scr{0}{j}) \leq 0$ and
      $\varphi^{j'}(v\scr{0}{j}) \geq \bar v^{j'}$ for $j' \neq j$. Combining
      with \autoref{lem:ubvv}, we further have that $\varphi^{j'}(v\scr{0}{j}) =
      \bar v^{j'}$ for $j' \neq j$.

      Then consider the two cases where $v_i$ is either $v$ or $v^{(-j)}$.
      \begin{itemize}
        \item In both cases, the allocations of item $j' \neq j$ won't change,
              because either item $j'$ is always allocated to buyer $i'$ or
              always allocated uniformly at random.
        \item $v_i = v$. In this case, all the buyers except $i'$ have the same
              value and hence the same allocation. In particular, they will get
              $1 / (n - 1)$ of item $j$.
        \item $v_i = v^{(-j)}$. In this case, according to the assumption
              \eqref{eq:assump}, buyer $i$ has the highest (positive) virtual
              value on item $j$ and will be allocated the entire item. To ensure
              that buyer $i$ in this case won't have incentive to misreport her
              values as $v$, she will be charged $v\scr{i}{j}(n - 2) / (n - 1)$
              for the extra $(n - 2) / (n - 1)$ fraction of item $j$ comparing
              with the previous case.
      \end{itemize}
      Given the previous analysis, if buyer $i$ has any value $v'$ with
      ${v'}\scr{i}{j} > v\scr{i}{j}$, misreporting her value as $v$ is strictly
      dominanted by misreporting as $v^{(-j)}$. Due to the complementary
      slackness condition \eqref{eq:ds-slack},
      \begin{align*}
        \zeta^*(v', v) = 0,
      \end{align*}
      which implies that
      \begin{align*}
        \varphi^j(v) \geq v\scr{i}{j}.
      \end{align*}

      By the assumption \eqref{eq:assump} and \autoref{lem:ubvv}, we get a
      contradiction:
      \begin{align*}
        v\scr{i}{j} \leq \varphi^j(v) < \varphi^j(v^{(-j)}) \leq v\scr{i}{j}.
      \end{align*}
    \end{proofof}

  \section{Missing Proofs}

    \begin{proofof}{\autoref{lem:regular-exist}}
      For condition \eqref{eq:regular-virtual}, if $\mu_{-i}(\v_{-i}) = 0$, then
      $\mu(\v) = 0$, and we can simply let
      \begin{align*}
        \forall v_i, v'_i \in \V_i,~
          \zeta^*_i(v_i, v'_i; \v_{-i}) = \eta^*_i(\v) = 0.
      \end{align*}
      By doing so, $\delta^*$ is still a feasible solution to the dual program,
      and the objective value does not change. Hence the optimality is
      preserved. Moreover,
      \begin{align*}
        {\phi^*}\scr{i}{j}(\v) = \eta^*_{i}(\v)v\scr{i}{j} + \sum_{v'_{i}}\left(
          \zeta^*_{i}(v_{i}, v'_{i}; \v_{-i}) v\scr{i}{j}
            - \zeta^*_{i}(v'_{i}, v_{i}; \v_{-i}){v'}\scr{i}{j}\right) = 0,
      \end{align*}
      as desired by \eqref{eq:regular-virtual}.

      In addition, when $v_i = \Zero$, $\mu_i(\Zero) = 0$, and
      $\mu_{-i}(\v_{-i}) > 0$,
      \begin{align*}
        {\phi^*}\scr{i}{j}(\v) = \eta^*_{i}(\v)v\scr{i}{j} + \sum_{v'_{i}}\left(
          \zeta^*_{i}(v_{i}, v'_{i}; \v_{-i}) v\scr{i}{j}
            - \zeta^*_{i}(v'_{i}, v_{i}; \v_{-i}){v'}\scr{i}{j}\right)
          = - \sum_{v'_{i}} \zeta^*_{i}(v'_{i}, v_{i}; \v_{-i}){v'}\scr{i}{j}
          \leq 0.
      \end{align*}

      For condition \eqref{eq:regular-source} and \eqref{eq:regular-trans}, we
      do the following changes:
      \begin{itemize}
        \item $\forall \v \in \V$, $i \in [n]$, $v_i \neq \Zero$,
              \begin{gather*}
                \tld \eta^*_i(\v) = 0 \quad \text{and} \quad
                \tld \eta^*_i(\Zero, \v_{-i}) = \mu_{-i}(\v_{-i}),  \\
                \tld \zeta^*_i(v_i, \Zero; \v_{-i})
                  = \zeta^*_i(v_i, \Zero; \v_{-i}) + \eta^*_i(\v)
                  \quad \text{and} \quad
                \tld \zeta^*_i(\Zero, v_i; \v_{-i})
                  = \zeta^*_i(\Zero, v_i; \v_{-i}) + \beta^*_i(\v).
              \end{gather*}
        \item All others remain the same.
      \end{itemize}

      Then we verify one by one that (i) the constructed $\tld \delta^*$ is
      still feasible ($\tld \delta^* \in \D$), (ii) regularization condition
      \eqref{eq:regular-source} and \eqref{eq:regular-trans} are satisfied, and
      (iii) the objective remains the same ($\obj(\tld \delta^*) =
      \obj(\delta^*)$).
      \begin{enumerate}
        \item $\tld \delta^* \in \D$: clearly, all the variables in $\tld
              \delta^*$ are still non-negative.

              Then we show that for $v_i \neq \Zero$,
              $\tld {\phi^*}\scr{i}{j}(\v) = {\phi^*}\scr{i}{j}(\v)$:
              \begin{align*}
                \tld {\phi^*}\scr{i}{j}(\v) &=
                  \tld \eta^*_i(\v)v\scr{i}{j} + \sum_{v'_i}\left(
                    \tld \zeta^*_{i}(v_i, v'_i; \v_{-i}) v\scr{i}{j}
                    - \tld \zeta^*_{i}(v'_i, v_i; \v_{-i}){v'}\scr{i}{j}
                  \right)  \\
                  &= (\tld \eta^*_i(\v)
                      + \tld \zeta^*_{i}(v_i, \Zero; \v_{-i}))v\scr{i}{j} +
                      \sum_{v'_i \neq \Zero}\left(
                        \tld \zeta^*_{i}(v_i, v'_i; \v_{-i}) v\scr{i}{j}
                        - \tld \zeta^*_{i}(v'_i, v_i; \v_{-i}){v'}\scr{i}{j}
                      \right) - \tld \zeta^*_{i}(\Zero, v_i; \v_{-i}) 0  \\
                  &= (\eta^*_i(\v) + \zeta^*_{i}(v_i, \Zero; \v_{-i}))
                      v\scr{i}{j} + \sum_{v'_i \neq \Zero}\left(
                        \zeta^*_{i}(v_i, v'_i; \v_{-i}) v\scr{i}{j}
                        - \zeta^*_{i}(v'_i, v_i; \v_{-i}){v'}\scr{i}{j}
                      \right) - \zeta^*_{i}(\Zero, v_i; \v_{-i}) 0  \\
                  &= {\phi^*}\scr{i}{j}(\v);
              \end{align*}
              $\tld {\phi^*}\scr{i}{j}(\Zero, \v_{-i}) \leq
                {\phi^*}\scr{i}{j}(\Zero, \v_{-i})$:
              \begin{align*}
                \tld {\phi^*}\scr{i}{j}(\Zero, \v_{-i}) &=
                  -\sum_{v'_i} \tld\zeta^*_i(v'_i,\Zero;\v_{-i}){v'}\scr{i}{j}
                  = - \sum_{v'_i} \zeta^*_i(v'_i, \Zero; \v_{-i}) {v'}\scr{i}{j}
                    - \sum_{v'_i} \eta^*_i(v'_i, \v_{-i}) {v'}\scr{i}{j}
                  \leq {\phi^*}\scr{i}{j}(\Zero, \v_{-i});
              \end{align*}
              for $v_i \neq \Zero$, $\tld \psi^*_i(\v) = \mu(\v)$:
              \begin{align*}
                \tld \psi^*_i(\v) &=
                  \tld \eta^*_i(\v) + \sum_{v'_i}\left(
                    \tld \zeta^*_{i}(v_i, v'_i; \v_{-i})
                    - \tld \zeta^*_{i}(v'_i, v_i; \v_{-i})\right)  \\
                  &= \tld \eta^*_i(\v)
                      + \tld \zeta^*_{i}(v_i, \Zero; \v_{-i}) +
                      \sum_{v'_i \neq \Zero}\left(
                        \tld \zeta^*_{i}(v_i, v'_i; \v_{-i})
                        - \tld \zeta^*_{i}(v'_i, v_i; \v_{-i})
                      \right) - \tld \zeta^*_{i}(\Zero, v_i; \v_{-i})  \\
                  &= \eta^*_i(\v) + \zeta^*_{i}(v_i, \Zero; \v_{-i})
                      + \sum_{v'_i \neq \Zero}\left(
                        \zeta^*_{i}(v_i, v'_i; \v_{-i})
                        - \zeta^*_{i}(v'_i, v_i; \v_{-i})
                      \right) - \zeta^*_{i}(\Zero, v_i; \v_{-i}) - \beta^*_i(\v)
                        \\
                  &= \psi^*_i(\v) - \beta^*_i(\v) = \mu(\v);
              \end{align*}
              and finally if $\mu_{-i}(\v_{-i}) = 0$, by previous constructions,
              $\tld \psi^*_i(\Zero, \v_{-i}) = 0 = \mu_{-i}(\v_{-i})$;
              otherwise,
              \begin{align*}
                \tld \psi^*_i(\Zero, \v_{-i}) =\;&
                  \tld \eta^*_i(\Zero, \v_{-i}) + \sum_{v'_i}\left(
                    \tld \zeta^*_{i}(\Zero, v'_i; \v_{-i})
                    - \tld \zeta^*_{i}(v'_i, \Zero; \v_{-i})\right)  \\
                  =\;& \mu_{-i}(\v_{-i}) + \sum_{v'_i}\left(
                    \zeta^*_i(\Zero, v'_i; \v_{-i}) + \beta^*_i(v'_i, \v_{-i})
                    - \zeta^*_i(v'_i, \Zero; \v_{-i}) - \eta^*_i(v'_i, \v_{-i})
                    \right)  \\
                  =\;& \mu_{-i}(\v_{-i}) + \sum_{v'_i}\left(
                    \zeta^*_i(\Zero, v'_i; \v_{-i})
                    - \zeta^*_i(v'_i, \Zero; \v_{-i})\right)  \\
                   & + \sum_{v_i \neq \Zero} \left(\sum_{v'_i} \left(
                      \zeta^*_i(v_i, v'_i; \v_{-i})
                      - \zeta^*_i(v'_i, v_i; \v_{-i})\right) - \mu(\v)\right)
                      \\
                  =\;& \mu_{-i}(\v_{-i}) - \sum_{v_i \neq \Zero} \mu(\v)
                     = \mu(\Zero, \v_{-i}).
              \end{align*}
        \item Regularization condition \eqref{eq:regular-source}: directly
              implied by construction, and \eqref{eq:regular-trans}: implied by
              $\tld \psi^*_i(\v) = \mu(\v)$, which is proved above.
        \item The objective remains the same because we did not change variables
              ${\xi^*}\sci{j}(\v)$ at all.
      \end{enumerate}
    \end{proofof}

    \begin{proofof}{\autoref{lem:bayes-regular-exist}}
      Similar to the proof of \autoref{lem:regular-exist}, omitted.
    \end{proofof}

    \begin{proofof}{\autoref{lem:ubvv}}
      Omitted.
    \end{proofof}

\section{Interpretations for the Duals}

  In this section, we provided some interpretations of the dual problems of the
  revenue maximization problem under the dominant-strategy implementation and
  the Bayesian implementation. In particular, they can be thought as an extended
  version of the interpretation by
  \cite{daskalakis2013mechanism,daskalakis2015strong}.

  To be added.

\section{Recover Myerson's Result for the Single-Item Setting}

  To be added.

\section{Future Work}

  We plan to entend our results for \autoref{sec:iid} to either (i) independent
  but non-identical cases, or (ii) continuous distribution cases.

\bibliography{virtual}
\bibliographystyle{named}

\clearpage

\appendix

\end{document}

%% file: header.tex
\usepackage{framed}
\usepackage{amsthm}
\usepackage{amsmath}
\usepackage{amssymb}
\usepackage{mathrsfs}
\usepackage{xspace}
\usepackage{booktabs}
\usepackage{aliascnt}
\usepackage{natbib}
\usepackage{balance}
\usepackage{enumitem}
\usepackage{libertine}
\usepackage{zi4}
\usepackage{newtxmath}

\usepackage[pdfauthor={Song Zuo},pdftitle={Generalizing Virtual Values to Multidimensional Auctions: a Non-Myersonian Approach},pagebackref=true,colorlinks]{hyperref}
\hypersetup{linkcolor=red,filecolor=red,citecolor=red,urlcolor=red}

\newtheorem{theorem}{Theorem}[section]
\newtheorem*{theorem*}{Theorem}

\newaliascnt{lemma}{theorem}
\newaliascnt{assumption}{theorem}
\newaliascnt{proposition}{theorem}
\newaliascnt{corollary}{theorem}
\newaliascnt{claim}{theorem}
\newaliascnt{observation}{theorem}
\newaliascnt{definition}{theorem}
\newaliascnt{fact}{theorem}
\newaliascnt{statement}{theorem}
\newaliascnt{mechanism}{theorem}

\newtheorem{lemma}[lemma]{Lemma}

\newtheorem{corollary}[corollary]{Corollary}
\newtheorem{definition}[definition]{Definition}

\aliascntresetthe{lemma}
\aliascntresetthe{assumption}
\aliascntresetthe{proposition}
\aliascntresetthe{corollary}
\aliascntresetthe{claim}
\aliascntresetthe{observation}
\aliascntresetthe{definition}
\aliascntresetthe{fact}
\aliascntresetthe{statement}
\aliascntresetthe{mechanism}

\newenvironment{proofof}[1]{\par{\noindent \it Proof of #1.}}{\qed\par}

\newcommand{\argmax}{\operatornamewithlimits{argmax}}

\newcommand\R{\mathbb{R}}

\newcommand\BRev{\textsc{BRev}}
\newcommand\DRev{\textsc{DRev}}
\newcommand\SRev{\textsc{SRev}}

\DeclareMathOperator*{\E}{\mathbb E}

\newcommand{\Infer}{\Longrightarrow}
\newcommand{\Reduce}{\Longleftarrow}

\newcommand{\V}{\mathcal{V}}
\newcommand{\x}{\boldsymbol{x}}

\newcommand{\Zero}{\boldsymbol{0}}
\newcommand{\One}{\boldsymbol{1}}
\renewcommand{\v}{\boldsymbol{v}}
\newcommand{\p}{\boldsymbol{p}}

\newcommand{\obj}{\mathrm{obj}}

\renewcommand{\P}{\mathcal{P}}
\newcommand{\D}{\mathcal{D}}

\newcommand{\scb}[1]{_{#1}}
\newcommand{\sci}[1]{^{#1}}
\newcommand{\scr}[2]{\scb{#1}\sci{#2}}

%% file: virtual.bbl
\begin{thebibliography}{}

\bibitem[\protect\citeauthoryear{Cai \bgroup \em et al.\egroup
  }{2012}]{cai2012optimal}
Yang Cai, Constantinos Daskalakis, and S~Matthew Weinberg.
\newblock Optimal multi-dimensional mechanism design: Reducing revenue to
  welfare maximization.
\newblock In {\em Foundations of Computer Science (FOCS), 2012 IEEE 53rd Annual
  Symposium on}, pages 130--139. IEEE, 2012.

\bibitem[\protect\citeauthoryear{Daskalakis \bgroup \em et al.\egroup
  }{2013}]{daskalakis2013mechanism}
Constantinos Daskalakis, Alan Deckelbaum, and Christos Tzamos.
\newblock Mechanism design via optimal transport.
\newblock In {\em Proceedings of the fourteenth ACM conference on Electronic
  commerce}, pages 269--286. ACM, 2013.

\bibitem[\protect\citeauthoryear{Daskalakis \bgroup \em et al.\egroup
  }{2015}]{daskalakis2015strong}
Constantinos Daskalakis, Alan Deckelbaum, and Christos Tzamos.
\newblock Strong duality for a multiple-good monopolist.
\newblock In {\em Proceedings of the Sixteenth ACM Conference on Economics and
  Computation}, pages 449--450. ACM, 2015.

\bibitem[\protect\citeauthoryear{Dobzinski \bgroup \em et al.\egroup
  }{2011}]{DBLP:conf/stoc/DobzinskiFK11}
Shahar Dobzinski, Hu~Fu, and Robert~D. Kleinberg.
\newblock Optimal auctions with correlated bidders are easy.
\newblock In Lance Fortnow and Salil~P. Vadhan, editors, {\em Proceedings of
  the 43rd {ACM} Symposium on Theory of Computing, {STOC} 2011, San Jose, CA,
  USA, 6-8 June 2011}, pages 129--138. {ACM}, 2011.

\bibitem[\protect\citeauthoryear{Myerson}{1981}]{myerson1981optimal}
Roger~B Myerson.
\newblock Optimal auction design.
\newblock {\em Mathematics of operations research}, 6(1):58--73, 1981.

\bibitem[\protect\citeauthoryear{Yao}{2016}]{yao2016solutions}
Andrew Chi-Chih Yao.
\newblock On solutions for the maximum revenue multi-item auction under
  dominant-strategy and bayesian implementations.
\newblock {\em arXiv preprint arXiv:1607.03685}, 2016.

\end{thebibliography}
